# Thin amorphous molybdenum silicide superconducting shells around individual nanowires deposited via magnetron co-sputtering

*Luize Dipane[1*], Martins Zubkins[1], Gunta Kunakova[2], Eriks Dipans[1], Tom Yager[1,3], Boris Polyakov[1], Edgars Butanovs[1*]*

[1]Institute of Solid State Physics, University of Latvia, Kengaraga street 8, LV-1063, Riga, Latvia

[2]Institute of Chemical Physics, University of Latvia, Jelgavas street 1, LV-1004, Riga, Latvia

[3]Current Address: International Iberian Nanotechnology Laboratory, Avenida Mestre José Veiga, 4715-330 Braga, Portugal

*Corresponding authors: luize.dipane@cfi.lu.lv, edgars.butanovs@cfi.lu.lv

**Abstract**

Employing amorphous superconductors, such as Type-II molybdenum silicide (MoSi), instead of crystalline materials significantly simplifies the material deposition and scalable nanoscale prototyping, beneficial for quantum electronic and photonic device fabrication. In this work, deposition of amorphous superconductive MoSi thin films on flat and nanowire (NW) substrates was demonstrated via pulsed direct-current magnetron co-sputtering from molybdenum and silicon targets in an argon atmosphere. MoSi films were deposited on oxidized silicon wafers and $Ga_2O_3$ NWs with 6 nm $Al_2O_3$ insulating shell, grown around the NWs using atomic layer deposition, and studied using scanning and transmission electron microscopy, X-ray diffraction, and X-ray photoelectron spectroscopy. Four-point Cr/Au electrical contacts were defined on the thin films and on individual $Ga_2O_3$-$Al_2O_3$-MoSi core-shell NWs using lithography for low-temperature electrical measurements. By controlling the sputtering power of the targets and thus adjusting the molybdenum-to-silicon ratio in the MoSi films, their properties were optimized to achieve critical temperature $T_c$ of 7.25 K. Such superconducting shell NWs could provide new avenues for fundamental studies and interfacing with other materials for quantum device applications.

# 1. Introduction

Nanoscale quantum materials are at the forefront of contemporary physics and materials science offering a complex interplay between various quantum physical phenomena, low dimensionality, and novel device architectures, that promises a next-generation o electronic and photonic technologies with currently unavailable functionalities [1]. The research has been driven by recent developments in quantum information science and engineering, which includes quantum computing, communications, and sensing [2]. Many of the quantum electronic and photonic devices are based on superconducting materials and Josephson junctions. However, traditional Type-I superconductor-based devices breakdown under high magnetic fields, motivating research into superconducting materials with higher critical temperatures and fields [3].

One-dimensional (1D) nanomaterials, such as nanowires (NWs), provide several advantages over bulk or planar materials, such as size-tuneable properties and localization effects [4–6], facilitating the fundamental study of new phenomena [7]. NWs have also emerged as versatile building blocks for nanoscale devices due to their high crystalline quality and adaptable fabrication of heterostructures with various geometry, e.g., core-shell, for on-demand functionality [8–10]. Controllable epitaxial growth of hybrid semiconductor-superconductor NW heterostructures has enabled various applications in quantum electronics, such as topological and gate-controlled superconducting electronics [11], as well as studies of superconducting full-shell NWs provide new insights into fundamental quantum mechanics [12]. For example, Little-Parks oscillations have been observed in full-shell Al-around-InAs NWs by Vaitiekenas et al. [12], Vekris et al. [13] and Valentini et al. [14], as well as in $WS_2$ nanotubes [15]. Furthermore, NWs excel at single-photon detection (SPD) in various architectures: single semiconductor core-shell NW photogating SPDs at room temperature, heterostructured NW array avalanche diodes, and superconducting NW single-photon detectors (SNSPDs) [2,16–18]. While top-down fabricated meandered SNSPDs, typically using materials such as NbN and WSi thin films, represent one of the most mature technologies for single-photon applications today, single-photon detection based on bottom-up synthesized 1D nanomaterials remains an emerging and largely unexplored area [18,19].

Employing amorphous superconductors, such as molybdenum silicide MoSi, instead of crystalline ones significantly simplifies the material deposition and scalable nanoscale prototyping, especially for SNSPD fabrication, since low-temperature processes can be used and there is no epitaxial restriction for the substrate [20]. Amorphous MoSi exhibits type-II superconductivity, with a transition temperature ($T_c$) ranging from 6 K to 8.4 K, determined by its composition, thickness, and growth parameters [21–23]. The absence of crystalline grain boundaries in its amorphous structure reduces electron scattering and enhances performance, which is beneficial for sensitive detection applications [23,24]. These films are typically deposited by direct current (DC) or radio frequency (RF) magnetron co-sputtering from elemental targets under high vacuum, with careful tuning of the Mo:Si ratio, working pressure, and substrate temperature to achieve amorphous structure and optimal superconducting performance [21–23]. Atomic layer deposition (ALD) method successfully used for nitride superconductor, such as NbN [25], growth could potentially be extended to uniform and conformal deposition of superconducting MoSi films as well [26], if currently existing drawbacks of residual byproduct incorporation and limited stoichiometry control determined by precursor chemistry are addressed. The highest $T_c$ values, up to 8.4 K, are obtained for films with approximately 78-83% Mo content, while at higher Mo concentrations the material becomes unstable and tends to crystallize into $Mo_3Si$, resulting in deteriorated superconducting properties [23]. Due to its stable amorphous morphology, MoSi is widely used in SNSPDs in which its Type-II behaviour and high critical field are advantageous, enabling very high detection efficiency and operation under elevated bias currents [27–29]. Furthermore, MoSi detectors have been integrated into photonic platforms such as lithium niobate and silicon nitride, demonstrating their compatibility with integrated photonic circuits and suitability for scalable fabrication [30].

Motivated by the limited number of studies on superconducting full-shell NWs, especially with amorphous superconductor, in this work we deposited thin amorphous MoSi shell around a $Ga_2O_3$-$Al_2O_3$ NW core as a passive template using pulsed-DC magnetron co-sputtering and explored the structural and functional properties of these core-shell NW. Low-temperature electrical measurements of an individual MoSi-based core-shell NW revealed a robust superconducting transition at 7.3 K. Such superconducting shell NWs could provide new avenues for fundamental studies and interfacing with other materials for quantum photonic and electronic device applications.

## 2. Materials and Methods

The $Ga_2O_3$-$Al_2O_3$-MoSi core-shell NWs were produced in a multiple step process. Firstly, synthesis of $Ga_2O_3$ NWs, then, deposition of an amorphous $Al_2O_3$ layer using ALD for the insulation of the semiconductor $Ga_2O_3$ core, followed by co-sputtering of Mo and Si using pulsed-DC magnetron sputtering. $Ga_2O_3$ NWs were selected as a passive template for superconducting shell deposition, while $Al_2O_3$ interlayer electrically insulates the MoSi shell from the semiconducting $Ga_2O_3$ core [31,32]. $Ga_2O_3$ NWs were chosen because they satisfy the key practical requirements for the core-shell NW preparation but do not have a scientific justification: sufficient length, straight morphology with no kinks for successful device fabrication via optical lithography, and adequate spatial separation between the vertically freestanding NWs to avoid shadowing effects during MoSi sputter-deposition, therefore, other material NWs could also be used provided they meet the same criteria. The MoSi layer was deposited on Ga$_2$O$_3$ NW coated with Al$_2$O$_3$ as well as on oxidized silicon wafers as planar thin films for comparison with the core-shell NWs.

The Ga$_2$O$_3$ NWs were synthesized by atmospheric pressure chemical vapor deposition (CVD) using a horizontal three-zone quartz tube reactor (Carbolite TG3-12-60-600). A ceramic boat loaded with 0.15 g of Ga$_2$O$_3$ powder (99.99 %, Aldrich) was placed in the first zone, heated at 1010 °C. SiO$_2$/Si(100) wafers (SemiconductorWafer, Inc.) coated with 40 nm Au nanoparticles (ThermoScientific) served as substrates and were positioned in the second and third zones at 800 °C. The Au nanoparticles acted as catalysts for NW growth via the vapor-liquid-solid (VLS) mechanism. The reactor was supplied with an Ar/H$_2$ (5 %) gas flow at 200 standard cubic centimetres per minute (sccm). When the target temperatures in all zones were reached (1000 °C, 800 °C, and 800 °C, respectively), the system was held at temperature for 30 minutes to facilitate NW growth. The reactor was then cooled to room temperature under constant gas flow.

As-prepared Ga$_2$O$_3$ NWs were coated with an insulating amorphous Al$_2$O$_3$ layer using a Veeco ALD Savannah S100 reactor. The deposition was performed at 150 °C for 66 ALD cycles, resulting in an Al$_2$O$_3$ film uniformly around NWs, approximately 6 nm thick. Each cycle alternated between Trimethylaluminum (TMA) and H$_2$O as precursors, with nitrogen (N$_2$) as the inert carrier gas.

The MoSi films were deposited via pulsed-DC magnetron co-sputtering from 2 inches molybdenum (99.95 %) and 3 inches silicon (99.999%) targets operated at 80 kHz with 2 µs reverse time for Mo and 70 kHz with a 3 µs reverse time for Si, in an argon atmosphere. The base pressure prior to deposition was 9.3 x 10$^{-7}$ mbar, and the process pressure was maintained at 4 × 10$^{-3}$ mbar by a throttle valve in a constant argon (6N) flow of 25 sccm. The substrates were mounted on a rotating holder and were not heated intentionally. MoSi films were deposited both on oxidized silicon wafers and on the previously described Ga$_2$O$_3$ NWs coated with Al$_2$O$_3$. To prepare thin films, lift-off resist (LOR 3A, micro resist technology GmbH) and photoresist (AZ1518, Merck Performance Materials GmbH) were spin-coated on the Si/SiO$_2$, and soft-baked at 180 °C and 200 °C, respectively. An optical mask with a multiple size square pattern was aligned on the substrate using Mask aligner Suss MA6 and exposed to UV light. After developing, MoSi thin film was deposited and lift-off process was carried out. To achieve amorphous films, multiple deposition runs were performed, maintaining the sputtering power of Mo at 33 W while varying the sputtering power of Si between 40 W and 50 W to obtain different stoichiometries. MoSi was deposited on thin films and NWs under identical deposition parameters, resulting in thickness of 92 nm and 22 nm, respectively, with the reduced thickness on NWs due to their higher surface area. Subsequently, 2 nm of Si was deposited at 70 W to serve as a capping layer to prevent oxidation [27].

As-synthesised NW morphology and distribution density was studied using a scanning electron microscope (SEM, Lyra, Tescan) and their inner crystalline structure examined via transmission electron microscope (TEM) using a Tecnai GF20, FEI operated at an accelerating voltage of 200 kV. X-ray diffraction (XRD) analysis was performed using Rigaku MiniFlex 600 111 powder diffractometer with Bragg-Brentano θ-2θ geometry and a 600 W Cu anode X-ray tube 112 (Cu Kα line, λ = 1.5406 Å) to investigate the phase composition.

The chemical composition of the MoSi-based core-shell NWs and thin films was studied with X-ray photoelectron spectroscopy (XPS) measurements performed using ESCALAB Xi spectrometer (ThermoFisher). Al Kα X-ray tube with the energy of 1486 eV was used as an excitation source, the size of the analysed sample area was 650 μm x 100 μm and the angle between the analyser and the sample surface was 90°. An electron gun was used to perform charge compensation; however, no sputter-cleaning was performed prior the measurements. The base pressure during the spectra acquisition was better than 10$^{-5}$ Pa.

To perform electronic characterisation, four-terminal and thin-film devices were fabricated using a standard cleanroom photolithography microfabrication process. First, the individual $Ga_2O_3$-$Al_2O_3$-MoSi core-shell NWs were mechanically dry-transferred onto standard Si-$SiO_2$ wafers (300 nm thermal oxide, Si conductivity < 0.005 Ohm-cm, Biotain Crystal Co., Limited), subsequently, lift-off resist and photoresist were spin-coated onto the substrate. An optical mask with a microelectrode pattern was then aligned over a single NW (or thin film) using a Mask aligner Suss MA6 and exposed to UV light. After the development, Cr/Au (4/86 nm) electrodes were deposited via thermal evaporation, followed by lift-off process. The charge transport characteristics of the deposited films and NW devices were measured using a Physical Property Measurement System (PPMS Dynacool9T) with the resistivity option in a four-probe configuration. In total three NW and two thin film superconducting samples were successfully measured (see Table S1).

## 3. Results and discussion

SEM imaging was used to determine the morphology of the as-prepared $Ga_2O_3$-$Al_2O_3$-MoSi core-shell NWs. As shown in Figure 1a, the NWs are well-distributed ensuring minimal NW overlap thus minimizing shadowing during MoSi sputter-deposition. The NWs typically reach lengths of up to 80 micrometres. As shown in Figure 1b, the core-shell NWs are straight and smooth, with uniform diameters. To investigate the core-shell NW inner crystalline structure, TEM imaging of individual NWs was performed. In Figure 1c it is visible that the NW exhibit a well-defined multilayer architecture. The high-resolution inset confirms the presence of a highly-crystalline $Ga_2O_3$ core, an intermediate 6 nm thick $Al_2O_3$ layer, and an outer MoSi shell. Each layer appears continuous and conformal along the NWs axis, indicating uniform coating during the deposition processes. Depending on the individual NW placement on the substrate during sputter-deposition, some thickness variations between opposite NW sides were observed in few cases due to the shadowing effect. Selected-area electron diffraction (SAED) pattern (see Figure S1) revealed bright diffraction spots which were be attributed to the crystalline $Ga_2O_3$ core, while a diffused halo ring pattern, presumably from the amorphous shells, can also be observed.

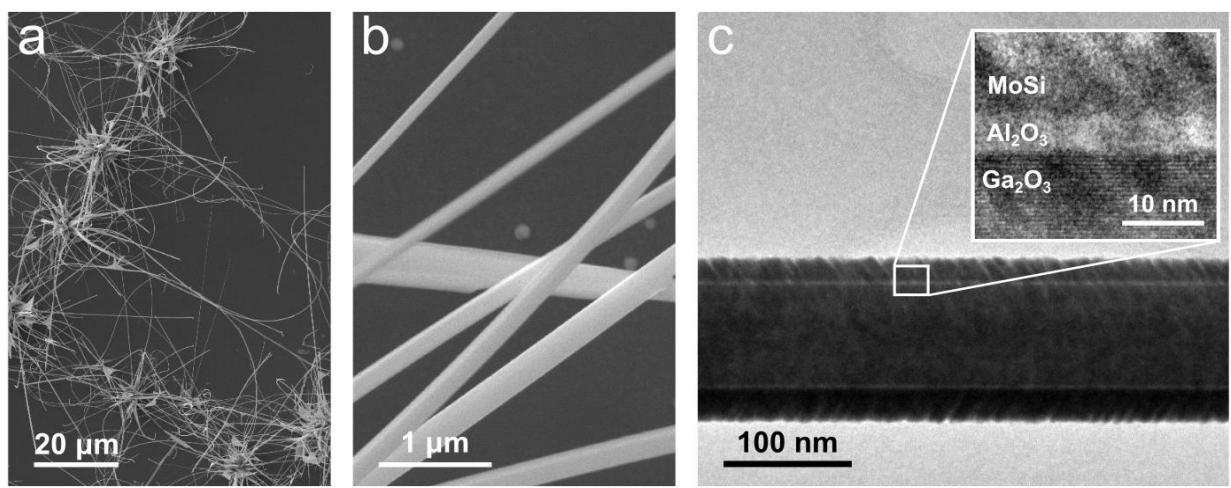

**Figure 1.** (a) Scanning electron microscope image showing the morphology of as-grown $Ga_2O_3$-$Al_2O_3$-MoSi core-shell NWs. (b) High-magnification scanning electron microscope image highlighting the smooth surface and uniform diameter of individual $Ga_2O_3$-$Al_2O_3$-MoSi NWs. (c) Transmission electron microscope image of the NW inner structure, revealing a multilayer configuration. The inset shows identifiable distinct layers of $Ga_2O_3$, $Al_2O_3$, and MoSi.

XRD patterns of MoSi thin films with varying Si content are shown in Figure 2. Only the peak from the Mo (110) is observed in the sample with the lowest Si content ($Mo_{0.79}Si_{0.21}$), consistent with ICDD-PDF No:00-004-0809. The Bragg peak at $2\theta \approx 33°$ is attributed to the Si(100) substrate and corresponds to the forbidden Si(200) reflection. Worth noting that the Si (100) substrate has an intense Si(400) peak at 69° that gives a strong background, therefore, for clarity purposes the patterns here are only showed until 60°. In the other samples with higher Si content, no peaks were detected, indicating that the films are amorphous. At lower Si concentrations, a crystalline phase, presumably Mo or $Mo_3Si$, emerges [22,23], while higher concentrations result in amorphous structures. The Si content in the thin films was controlled by adjusting the pulsed-DC power applied to the Si target during co-sputtering, while the Mo target power was kept constant at 33 W. Notably, the Si level required to maintain amorphous structures also depends on the sputtering power applied to the Mo target – at higher powers (e.g., > 40 W), crystalline phases can still form even at relatively high Si contents, as shown in Figure S2. Such film structure dependency (emergence of crystalline phases) on the sputtering power has been previously observed by Grotowski et al. [23], and is known to significantly degrade the film superconducting properties.

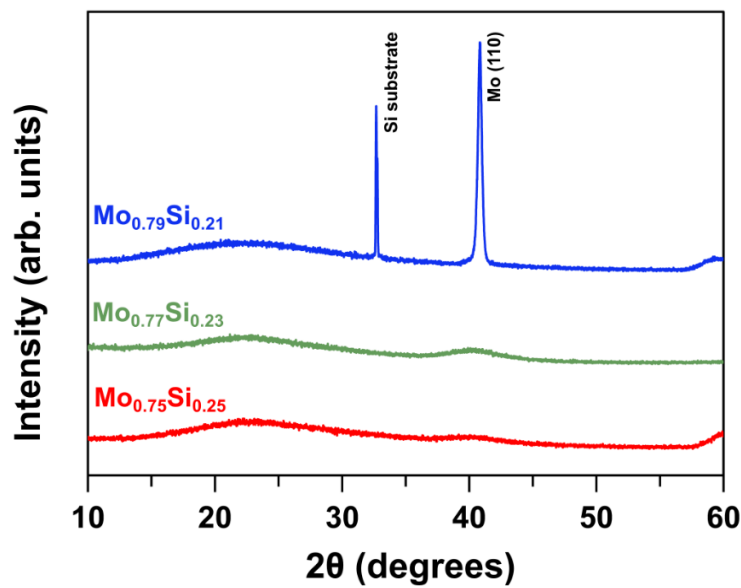

**Figure 2.** X-ray diffraction patterns of MoSi thin films with varying silicon content: $Mo_{0.79}Si_{0.21}$ (blue), $Mo_{0.77}Si_{0.23}$ (green), and $Mo_{0.75}Si_{0.25}$ (red). The patterns exhibit a broad amorphous background for all compositions, with crystalline peaks observed only in the $Mo_{0.79}Si_{0.21}$ sample, corresponding to the Mo(110) Bragg peak and Si substrate forbidden reflection.

The chemical states and stoichiometry of the elements in the surface of the as-prepared thin films and core-shell NWs were studied in detail using XPS. Survey spectra indicated the presence of Mo, Si and O elements, as well as carbon from the organic surface contaminants (see Figure 3(a)), and were used to provide information on the Mo:Si ratio for deposition optimization at different sputtering parameters. Stoichiometry calibration from the XPS data was performed on the uncapped samples (reference samples without the Si capping layer).

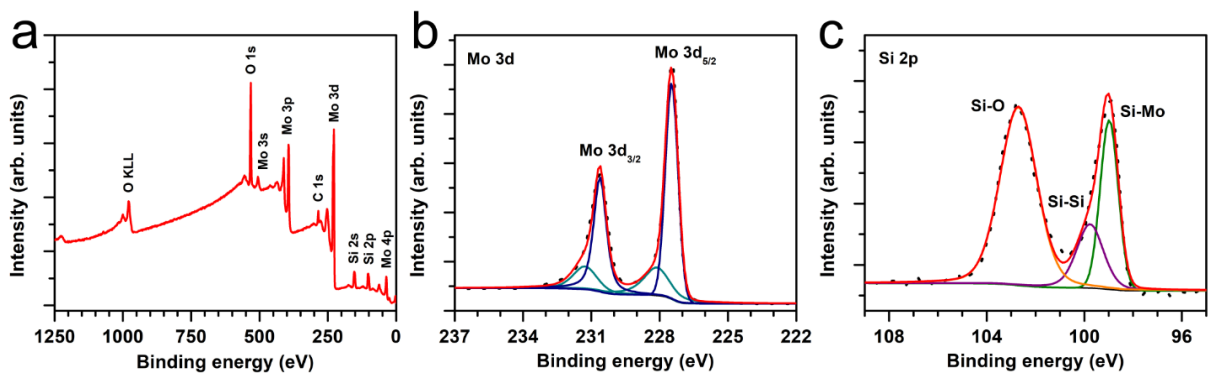

**Figure 3.** (a) XPS survey scan of the $Ga_2O_3$-$Al_2O_3$-MoSi core-shell NW arrays, and the corresponding high-resolution scans of (b) Mo 3d and (c) Si 2p regions.

High-resolution spectra of Mo and Si elements were acquired and calibrated relative to the adventitious C 1s peak at 284.8 eV, and are shown in Figure 3(b-c). Fitting of the Mo 3d scan revealed the presence of two doublets ($\Delta_{5/2-3/2}$ =3.1 eV), the most prominent peak matching $Mo^{4+}$ valence state at Mo $3d_{5/2}$ = 227.5 eV, similar to that of $MoSi_2$ [33,34], while the small contribution at Mo $3d_{5/2}$ = 228.1 eV could potentially be attributed to Mo-Mo bonding [35]. On the other hand, Si 2p scan indicated the presence of various Si bonding components for Si-Mo at 99.0 eV and Si-Si at 99.8 eV from the MoSi shell [34], and Si-O at 102.7 eV from the presumably oxidized 2 nm Si capping layer. The use of the capping layer for the protection of the amorphous MoSi shell in the final samples was motivated by a preliminary XPS study of uncapped core-shell NWs, shown in Figure S3. While an unprotected sample does not undergo significant oxidation in a couple of days after the synthesis (see Figure S3(a-c)), an uncapped sample kept in ambient conditions for 2 months showed significant oxidation, indicated by the emergence of distinct $MoO_3$ and $SiO_2$ components at higher binding energies [36] in the XPS spectra shown in Figure.S3(d-f). To maintain the integrity of the MoSi layers over longer periods of time, the sputtered MoSi layers were terminated by 2 nm of pure Si protective layers, which supposedly oxidizes after exposure to ambient conditions. Similar protection strategy was used by Reddy et al. [27].

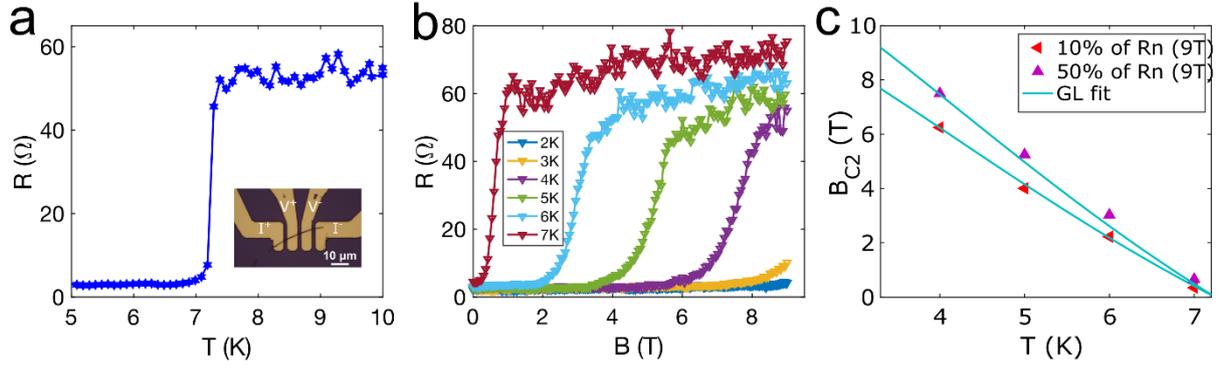

**Figure 4.** (a) Temperature-dependent resistance of a single $Ga_2O_3$-$Al_2O_3$-MoSi NW device highlighting the superconducting transition, with the resistance dropping sharply to near zero below the critical temperature of 7.25 K. The inset shows an optical microscope image of the fabricated device with metal contacts aligned on a single NW. The resistance is recorded in a four-probe configuration by measuring the inner distance between the V+/V- electrodes while applying excitation current to the I+/I- electrodes. (b) Resistance as a function of magnetic field recorded at various temperatures. The sample is aligned perpendicularly to the direction of the magnetic field. (c) Temperature dependence of the upper critical field $B_{c2}$ estimated using two criteria, corresponding to 10% and 50% of the normal state resistance value at a magnetic field of 9 T, respectively. The solid lines correspond to the fit of eq.(2).

Temperature-dependant resistivity measurements with and without applied external magnetic field were performed to understand the charge transport characteristics of the as-prepared individual core-shell NW devices, and were compared to those of thin films. Figure 4(a) shows the resistance recorded for an individual core-shell NW, in the absence of a magnetic field. The critical temperature $T_c$ is determined as the temperature at which the normal state resistance ($R_n$ = R at 10 K) decreases by 50%. For the $Ga_2O_3$-$Al_2O_3$ NW with a total core diameter of 252 nm and a MoSi-Si shell thickness of 24 nm, $T_c$ = 7.25 K (sample NW 16, Table S1), as expected of such $Mo_{0.77}Si_{0.23}$ stoichiometry [22]. For comparison, this value is only slightly lower than that measured for a 94 nm thick film ($T_c$ = 7.32 K, the same estimation criterion). In contrast, for another NW with a ~90 nm larger $Ga_2O_3$-$Al_2O_3$ core diameter and the same nominal MoSi-Si shell thickness, $T_c$ is reduced to 6.93 K, presumably indicating slight variations in stoichiometry or imperfect shell coverage over the larger diameter NW. At temperatures below $T_c$, a residual resistance of a few ohms is present, likely due to dissipation in the superconducting state and/or some remaining contact resistance from the presumably oxidized 2 nm Si capping layer barrier. Such residual resistance has been frequently observed for disordered superconducting NWs and their meanders [37,38] and can be attributed to phase-slip-governed processes and other mechanisms [39].

The NW device with the smallest core diameter is further analysed to evaluate the upper critical magnetic field $B_{c2}$, via the magnetic field induced transition to resistive state, using the magnetoresistance isotherms (Figure 4(b)). Here, $B_{c2}$ at various temperatures is determined by considering 10% and 50% of the normal state resistance [40]. The obtained $B_{c2}(T)$ can be used to estimate the Ginzburg-Landau coherence length at zero temperature $\xi(0)$, and $\xi(0) = \sqrt{\frac{\Phi_0}{2\pi B_{c2}(0)}}$ (eq.1), where $\Phi_0$ is the magnetic flux quantum. The critical field $B_{c2}$ at zero temperature could be determined either by a linear extrapolation of $B_{c2}(T)$ [24] or by applying the Werthamer-Helfand-Hohenberg (WHH) theory. The temperature dependence of $B_{c2}$ can also be empirically described as [41]: $B_{c2}(T) = B_{c2}(0)\frac{1-(T/T_c)^2}{1+(T/T_c)^2}$ (eq.2). Fitting the $B_{c2}(T)$ data (Figure 4(c)) using this equation yields different $B_{c2}(0)$ values for the two criteria of 10 and 50%, resulting $\xi(0)$ of 5.3 and 4.7 nm, respectively. The estimated $\xi(0)$ of approximately 5 nm ($T_c$ = 7.25 K, superconducting shell thickness of ~24 nm) is consistent with the values reported in literature for $Mo_xSi_{1-x}$ films [22]. This indicates that such core-shell NWs are a viable building block material for further studies on quantum photonic and electronic device fabrication and operation.

## 5. Conclusions

Fabrication of $Ga_2O_3$-$Al_2O_3$-MoSi core-shell NWs with a passive $Ga_2O_3$-$Al_2O_3$ NW core and amorphous MoSi superconducting shell was demonstrated, and robust superconducting transition at 7.25 K in an individual core-shell NW was achieved. The obtained superconducting transport properties of the developed full-shell NWs closely match those of planar thin films measured here and in other studies. Growth process optimization to obtain amorphous superconductive MoSi thin films indicated the importance of adjusting the molybdenum-to-silicon ratio, controlling the sputtering power, and optimizing the argon gas flow rate and pressure during the DC magnetron co-sputtering deposition. Four-point Cr/Au electrical contacts were defined on individual $Ga_2O_3$-$Al_2O_3$-MoSi core-shell NWs and thin film using optical lithography to perform low-temperature electrical measurements and evaluate the critical temperature $T_c$ and critical magnetic field $B_{c2}$ of the MoSi shell. While a core-shell NW with a total core diameter of 252 nm indicated properties very close to those of thin films, nearly 100 nm thicker NW exhibited reduced $T_c$ of 6.93 K, presumably indicating slight variations in stoichiometry or imperfect shell coverage over the larger diameter NW. To further advance the understanding of superconducting transport in these core-shell NWs, more detailed investigations should be carried out to identify dissipation mechanisms in the superconducting state and to accurately determine the upper critical field values, including those under magnetic fields aligned in parallel to the NW longitudinal axis. To the best of our knowledge, studies on amorphous superconducting shell NWs remain very limited, therefore, our work may pave the way for using such NWs for fundamental studies and interfacing with other materials for quantum device applications.

**Supplementary Materials:** Supporting information is available and contains data describing the structural, chemical, and electrical properties of MoSi samples, including measured geometries, resistance, and $T_c$ values (Table S1), SAED pattern of the $Ga_2O_3$-$Al_2O_3$-MoSi nanowire (Figure S1), XRD patterns showing crystallinity depending on sputtering power (Figure S2), and XPS spectra revealing surface oxidation over time (Figure S3).

**Funding:** This research was funded by the Latvian Council of Science project No. lzp-2022/1–0311. G.K and T.Y were supported by the Latvian Quantum Initiative under European Union Recovery and Resilience Facility project no. 2.3.1.1.i.0/1/22/I/CFLA/001.

**Acknowledgements:** -

**Author Contributions:** Conceptualization, E.B.; Methodology, E.B., M.Z., G.K., T.Y. Validation, L.D., G.K., E.B.; Investigation, L.D., G.K., E.D, B.P., E.B; Writing – Original Draft Preparation, L.D., E.B., G.K.; Writing – Review & Editing, L.D., E.B., B.P., G.K., M.Z., T.Y.; Visualization, L.D., G.K.; Supervision, E.B.; Project administration, E.B.

**Institutional Review Board Statement:** Not applicable.

**Informed Consent Statement:** Not applicable.

**Data Availability Statement:** The data supporting this study's findings are available from the corresponding author upon reasonable request.

**Conflicts of Interest:** The authors declare no conflict of interest.